\documentclass{article}
\usepackage{times}
\usepackage{graphicx} 
\usepackage{wrapfig}
\usepackage{subfigure} 
\usepackage{natbib}
\usepackage{algorithm}
\usepackage{algorithmic}
\usepackage{hyperref}

\usepackage{float}

\usepackage[accepted]{icml2015}

\icmltitlerunning{Driverseat: Crowdstrapping Learning Tasks for Autonomous Driving
}

\begin{document} 

\twocolumn[
\icmltitle{Driverseat: Crowdstrapping Learning Tasks for Autonomous Driving}

\icmlauthor{Pranav Rajpurkar}{pranavsr@cs.stanford.edu}
\icmlauthor{Toki Migimatsu}{takatoki@cs.stanford.edu}
\icmlauthor{Jeff Kiske}{jkiske@stanford.edu}
\icmladdress{Stanford University Computer Science Dept., 353 Serra Mall, Stanford, CA 94305 USA}
\icmlauthor{Royce Cheng-Yue}{rchengyue@gmail.com}
\icmladdress{3265 Falerno Way, San Jose, CA 95135 USA}
\icmlauthor{Sameep Tandon}{sameep@stanford.edu}
\icmlauthor{Tao Wang}{twangcat@stanford.edu}
\icmlauthor{Andrew Ng}{ang@cs.stanford.edu}
\icmladdress{Stanford University Computer Science Dept., 353 Serra Mall, Stanford, CA 94305 USA}

\icmlkeywords{Driverseat, crowdstrapping, human computation, lane detection, autonomous driving, machine learning, CrowdML, ICML}

\vskip 0.3in
]

\begin{abstract}

While emerging deep-learning systems have outclassed knowledge-based approaches in many tasks, their application to detection tasks for autonomous technologies remains an open field for scientific exploration. Broadly, there are two major developmental bottlenecks: the unavailability of comprehensively labeled datasets and of expressive evaluation strategies. Approaches for labeling datasets have relied on intensive hand-engineering, and strategies for evaluating learning systems have been unable to identify failure-case scenarios. Human intelligence offers an untapped approach for breaking through these bottlenecks. This paper introduces \textit{Driverseat}, a technology for embedding crowds around learning systems for autonomous driving. Driverseat utilizes crowd contributions for (a) collecting complex 3D labels and (b) tagging diverse scenarios for ready evaluation of learning systems. We demonstrate how Driverseat can \textit{crowdstrap} a convolutional neural network on the lane-detection task. More generally, crowdstrapping introduces a valuable paradigm for any technology that can benefit from leveraging the powerful combination of human and computer intelligence.
\end{abstract} 


\begin{figure}
\centering
\includegraphics[width=\columnwidth]{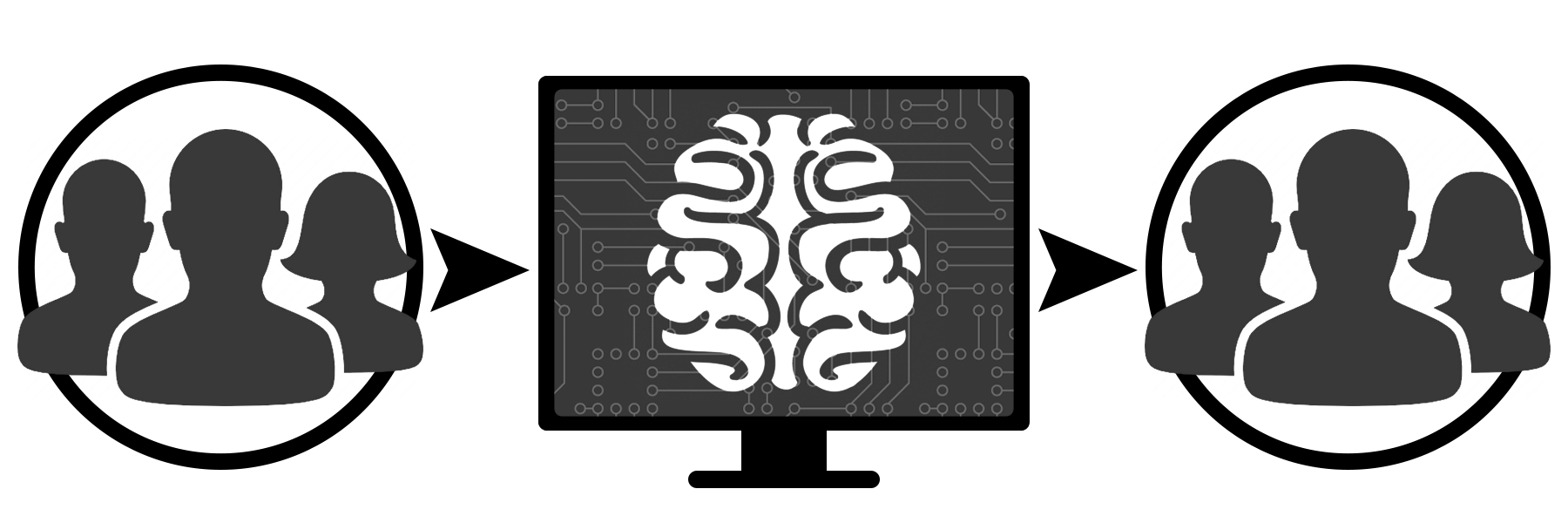}
\caption{Embedding the crowd around learning systems for both training and evaluation, we can leverage the powerful combination of human and computer intelligence.}
\label{fig:system}
\end{figure}

\section{Introduction}
\label{Introduction}

Autonomous driving is a mission critical technology of the future.
Road traffic accidents are responsible for nearly 1.3 million deaths globally each year, and are the leading cause of death among young people aged 15-29 \cite{WHO_Report}. In the quest to making roads accident-free, Advanced Driver Assistance Systems (ADAS) are making significant milestones: Adaptive Cruise Control systems adjust vehicle speed to maintain safe distance from vehicles ahead, and Automatic Braking systems react to imminent collisions \cite{markoff_sengupta_2013}. The next generation of ADAS will introduce features such as Traffic Jam Assistance, Traffic Light Detection, and Lane Change Assist, which enable a new suite of critical safety advancements \cite{road_progress}. Detection, or perception, is a key piece of the puzzle that the next generation of ADAS must solve \cite{ng_rss}: systems must perceive road boundaries, lane topologies, locations of other cars, pedestrians, signs, and miscellaneous obstacles. \cite{towards_fully}. Classic techniques to solve the detection problem have relied on intensive hand-engineering, and have been unable to capture the seemingly endless array of possibilities and conditions encountered on the road.

Deep learning systems represent an alternative approach \cite{deep_road}. Powered by large network architectures and fueled by the emergence of comprehensively labeled datasets \cite{ng_talk}, deep learning systems have made significant leaps in visual object recognition \cite{imagenet}, speech \cite{deep_speech}, and language understanding tasks \cite{socher_2013}. In the context of autonomous driving, progress for learning systems is bottlenecked by the unavailability of both comprehensively labeled datasets and measures to evaluate performance robustness.

On one end of the learning pipeline, approaches to labeling datasets for detection tasks in autonomous driving have focused on hand-engineering \cite{auto_extraction} or on synthetic generation of labels \cite{alvinn}. These approaches are (a) unable to capture the diversity and complexity of labels on roads (b) require months of heavy fine-tuning, and (c) are task-specific: an automated car-labeling system is not an effective lane-labeling system.

On the other end of the learning pipeline, approaches to evaluating the performance of learning systems for detection tasks have not been comprehensive \cite{kitti}. While metrics for general system performance are useful, they give little insight into system performance in adverse lighting, road, and weather conditions. In building a production-ready system for autonomous driving, it is vital to quantify and qualify its robustness under a slew of different road, lighting and weather scenarios.

Human intelligence offers an untapped approach for both ends of the pipeline. Driving is a complex, yet quotidian, task for people. With experience in the driver's seat, people have good intuition about road structures, car motion, lane topologies, and tricky environments. We hypothesize that we can integrate the experience of people to \textit{crowdstrap} learning systems for autonomous driving.

As a step towards integrating human expertise into reliable autonomous driving, we present \textit{Driverseat}, a web system that harnesses crowd contributions to bootstrap both the training and evaluation for learning systems. Two fundamental building blocks constitute Driverseat's architecture:

\begin{enumerate}
  \item \textbf{RoadEdit} leverages crowd contributions to train learning systems. It provides an toolbox for complex labeling tasks that are hard to automatically annotate.
  \item \textbf{TagEval} utilizes human intelligence to evaluate learning systems. It exposes a framework to tag diverse road scenarios, on which the performance of learning systems can be evaluated.
\end{enumerate}

The overarching contribution of the work is the idea of embedding human intelligence around learning systems for reliable autonomous driving. Beyond fueling the performance of learning systems, Driverseat is valuable in gaining intuition about the strengths and shortcomings of a learning system. For example, an early iteration of our detection system trained on data from California highways, which run North-South, performed poorly when facing the sun. In enabling such learnings, Driverseat guides research direction, especially important in building the perfect-performance systems expected for autonomous driving.

As a concrete demonstration of crowdstrapping learning tasks for autonomous driving, we focus on the lane detection task, an unfinished, yet vital, milestone for autonomous systems of the future. The lane detection task involves understanding the topology of the lanes around the car. It has been well investigated in the simple case, where the task is to detect the boundaries of the lane being driven in, also called the \textit{ego-lane}, for a short distance ahead \cite{on_the_road}. Approaches have relied on sophisticated modelling of road and motion \cite{gold, lane_cheng, lane_wang}, yet have not been able to accurately model lane topologies in their full complexity.

In the remainder of this paper, we (a) describe the engineering and design aspects of Driverseat, (b) detail how machine automation can be leveraged to seed the crowdlabeling task, and (c) demonstrate Driverseat's capability to crowdstrap a convolutional neural network for the lane detection task.

While the work uses the lane detection task in autonomous driving as an example, the idea of crowdstrapping learning systems introduces a valuable paradigm for any technologies that can benefit from leveraging the powerful combination of human and computer intelligence. Autonomous driving is a particularly powerful application, as we contribute to a future in which our lives are easier, technologies smarter, and world safer.

\section{Driverseat: Engineering and Design}
\label{Methodology}

Driverseat is a web system that utilizes human intelligence for (a) collecting complex labels and (b) identifying scenario-specific weaknesses in learning systems. The modules for achieving those goals are called RoadEdit and TagEval, respectively.

The significant technical contribution of Driverseat is that it introduces a 3D web interface for crowd-interaction. Traditional labeling interfaces exploit only 2D labeling \cite{label_me, image_net}. While 2D labeling is well suited for many tasks, complex labeling tasks in autonomous driving demand a more sophisticated 3D interaction interface. Some of the key shortcomings of 2D labeling for a task such as lane-labeling are that (a) depth information of the points is not easily visible or modifiable (b) points further from the car are hard to label accurately, (c) segments need to be labeled redundantly when multiple camera images shared the same stretch of the road, (d) lanes occluded by cars are difficult to label, and (e) piecewise-linear segments in 2D cannot capture sharp road-curves. Driverseat is designed to overcome the shortcomings of the 2D lane-labeling system, making labeling more expressive.

\subsection{RoadEdit Labeling}

While some labeling tasks involve little more than drawing a bounding box around an object in an image, other labeling tasks are more complex \cite{complex_crowd}. In autonomous driving, the key requirement of depth information heightens the complexity of labeling tasks. While there are techniques for automatically estimating depth maps \cite{automatic_3D}, implementation of such techniques requires a significant engineering effort. Driverseat's RoadEdit provides a simple alternative, with an interface that empowers the crowd to take over complex 3D labeling tasks.

\begin{figure*}
\centering
\includegraphics[resolution=144]{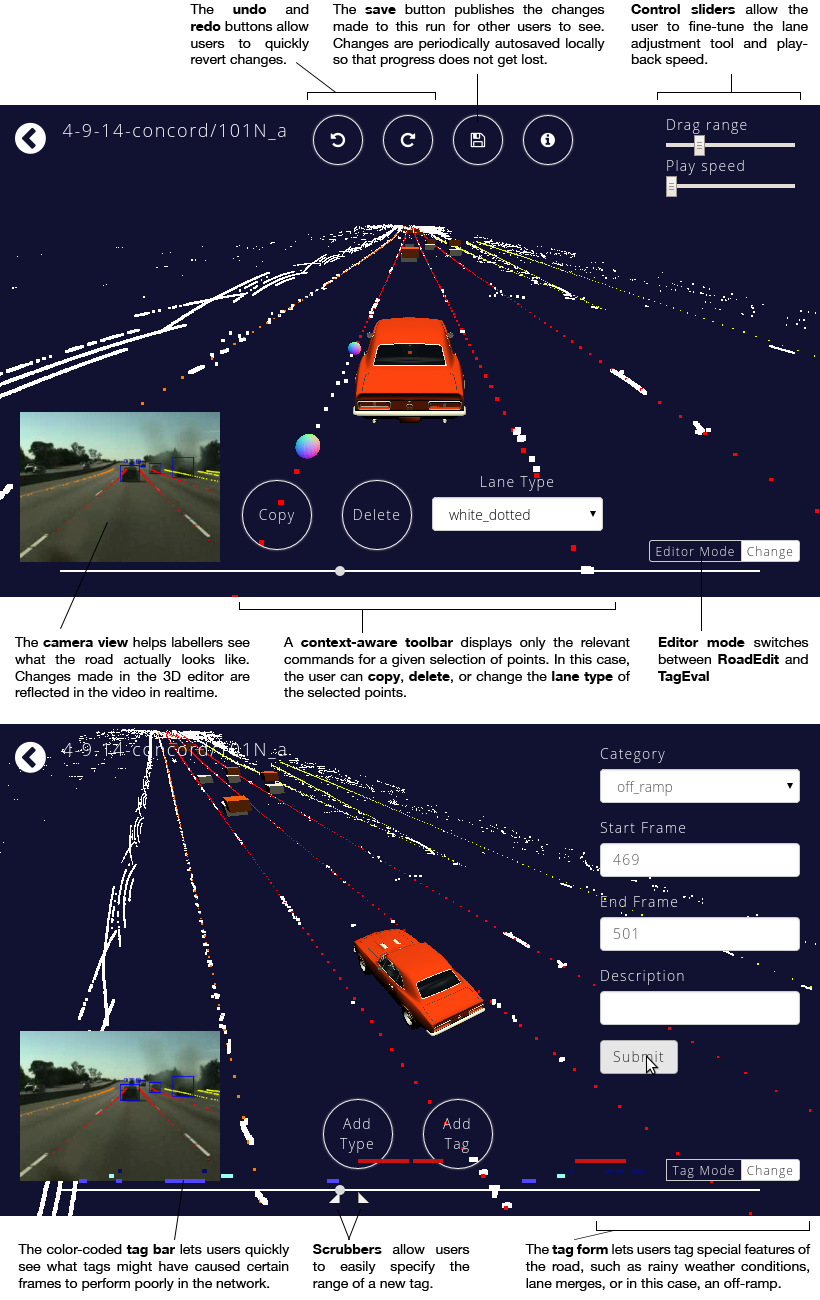}
\caption{The Driverseat building blocks: RoadEdit (above)  provides an toolbox for complex labeling tasks that are hard to automatically annotate, and TagEval (below) exposes  a  framework  to  tag  diverse road scenarios.}
\label{fig:interface}
\end{figure*}

We demonstrate RoadEdit in the concrete context of lane-labeling, one of the most challenging labeling tasks for autonomous driving. The user goal of RoadEdit is to have the colored lane boundaries run through white lane-paint points in 3D map (shown in figure \ref{fig:interface}). The user also assigns each lane, or lane-segment, a particular type, such as \textit{white dotted} or \textit{yellow solid}. Each type corresponds to a particular lane color.

Lane-labeling in RoadEdit enables capture of a variety of lane structures: lanes arbitrarily curve, merge with each other, split into several paths (such as a highway exit or the fan out of an urban arterial into turning lanes at an intersection) or end abruptly (at the threshold of an intersection). 

To capture these complex lane topologies, RoadEdit provides users with a variety of operations in their toolbox. Imprecise Lane boundaries can be dragged to ``line up'' on top of the lane markings. The drag range parameter controls the range of points affected by the drag, allowing for both short-range, and medium-range corrections. In addition to corrections, the user can use the \textit{fork} operation to model splits, \textit{append} to extend a lane, \textit{delete} and \textit{join} to model merges and abrupt ends. The combination of these operations can capture lane topologies of high complexity, which are unable to be be extracted automatically.

Although the RoadEdit toolbox is lane-labeling specific, the techniques applied generalize to complex 3D labeling in and beyond the context of autonomous driving. For instance, any 3D labelling system must tackle the user-interaction challenge that arises for 3D object manipulation on the 2D screen. For an object in 3D space, there are 6 degrees of freedom: 3 translational, and 3 rotational. Controlling all 6 degrees of freedom of an object is a nontrivial interaction task. Driverseat implements the well investigated workaround of introducing constraints on the control to limit the degrees of freedom \cite{virtual}. In the context of lane-labelling, when a lane is dragged, lane points are restricted to move only on the ground plane, restricting lane movement to two degrees of freedom. Another generalizable technique that RoadEdit leverages is the embedding of gamification in the labeling task, allowing people to label while enjoying themselves \cite{game}. Driverseat gamifies labelling by simulating the car's drive on the road. The virtual car moves through the the virtual environment, following the path of the data-collection vehicle. By combining simulation with labeling, we make the labeling task interactive and intuitive.

Another facet of RoadEdit that extends to other labelling tasks is its crowd-programming pattern, which follows a fix and verify strategy \cite{soylent}. The dataset is partitioned into individual \textit{drives}, where each drive consists of approximately 10 minute-long segments called \textit{runs}. Each run is marked by a \textit{marker-annotator} and verified by a \textit{verifier-annotator}. For quality control, we (a) employ  \textit{Expert Review} \cite{human_comp}, in which an annotated run is endorsed by an expert, and (b) use an expert to train and mentor a first-time annotator. Because Driverseat is exposed as a a web interface, trained annotators can perform labeling tasks remotely, and build on each other's progress. While this brings scalability to labelling, it poses the challenge of aggregating user contributions. To overcome this, RoadEdit implements a rudimentary version control system that enables undo/redo operations within and across labeling sessions.

\subsection{TagEval Evaluation}
Evaluation of learning systems rarely gives insight into where systems succeed and fail. This insight would be valuable in guiding research direction, especially in the context of autonomous driving, for which near perfect performance is critical. If systems are able to recognize particularly precarious scenarios for which the confidence of the detections are low, the driver can be preemptively alerted.

TagEval is Driverseat's scenario tagging interface used to tag road-conditions (splits, merges, bridges), lighting conditions (shadows, sun-facing), and weather conditions (rainy, snowy) on relevant road segments. These tags can then be used to evaluate how learning systems perform under these various scenarios.

The interface for TagEval is simple: labelers use a combination of the virtual 3D environment and the 2D camera images to tag road-segment conditions (shown in figure \ref{fig:interface}) . The tagging process involves selecting a start and end frame, and choosing a new/existing category associated with the tag. User access to multiple streams of information is beneficial: while the virtual 3D environment gives better clues about road and lane semantics, the 2D images give information about the environmental conditions.

\section{Seeding Human Computation With Automation}

Driverseat motivates the use of human computation in coordination with computer intelligence to solve complex problems \cite{street}. Having described how people can be used to label data for and evaluate learning systems, we now describe how computer automation can be used to supplement the human annotation task.

Lane-labeling is among a set of tasks that is complex for automatic labeling and time-consuming for human labeling. We describe how we can bootstrap automatic labeling to generate an initial ground-truth estimate, and ease the manual labeling task so that annotators only have to validate and make corrections to labels in complex scenarios \cite{soccer}.

To generate an initial ground-truth estimate, we adopt a two-step engineering methodology. Firstly, we extract the left and right-lane boundaries of the ego-lane. Secondly, we extend the ego-lane estimates to multiple lanes.

\begin{figure}
\centering
\includegraphics[width=\columnwidth]{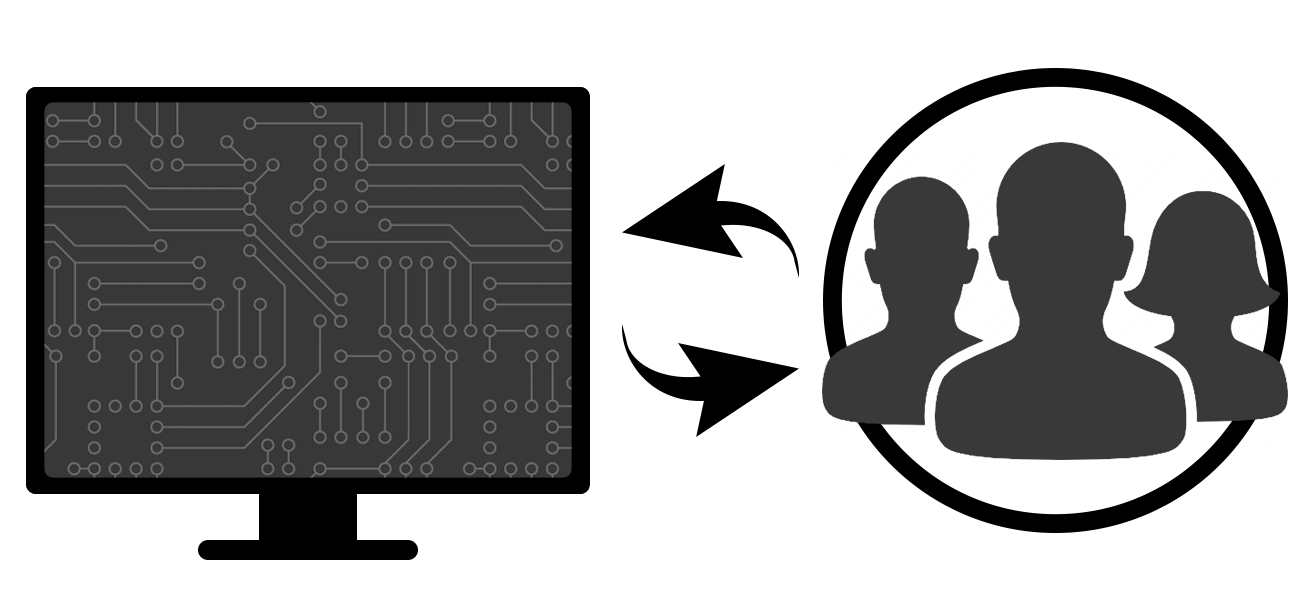}
\caption{Driverseat leverages a symbiosis between computers and people: machine annotations bootstrap human computation, which in turn bootstrap machine learning systems.}
\label{fig:computer-seed}
\end{figure}

For ego-lane boundary generation, we leverage information from the 3D road-maps and from data collection drives, during which lane changes are not performed by the driver. Thus, the GPS trajectory of the research vehicle runs within the ego-lane boundaries. We thus filter road points belonging to the ego-lane boundaries by keeping points that are within a lane-width distance from the GPS track. We further discriminate the points belonging to the left boundary from those belonging to the right boundary by using the sign of the lateral distance. After obtaining the points belonging to the left and right boundaries, we fit a piecewise linear curve to each boundary.

\begin{figure}
\centering
\includegraphics[width=\columnwidth]{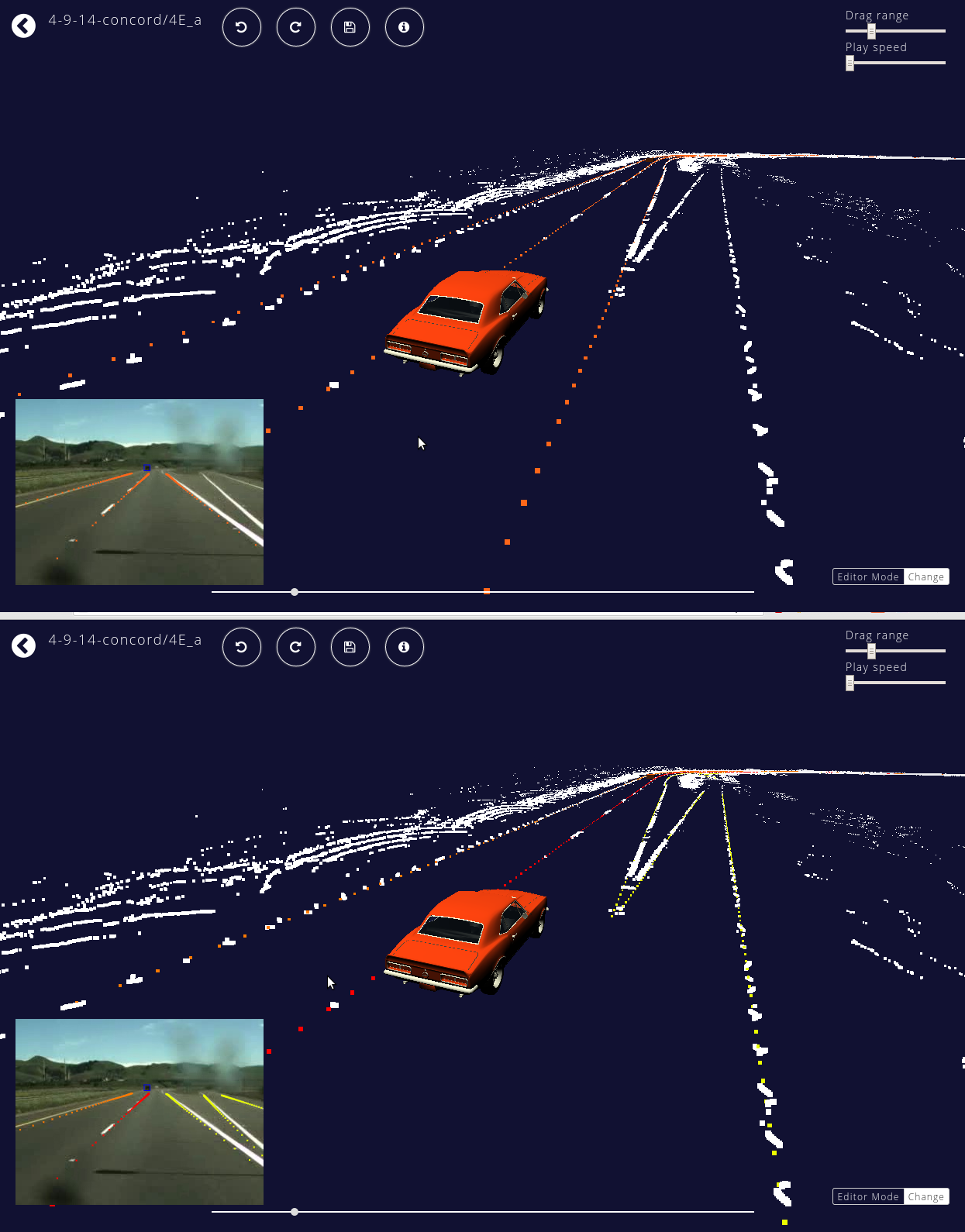}
\caption{(a) The automatic labeling generates decent results for the ego-lane, but is unable to capture complex lane topologies such as splits. (b) The crowd-workers, seeded by automatic labeling, can capture these complexities.}
\label{fig:before_after}
\end{figure}

Because ground points away from the ego-lane are sparse, the technique for ego-lane generation does not generalize well in the multilane case. To generate estimates for multiple lanes, we make the simplifying assumption that neighboring lanes follow a very similar structure to the ego-lane. We can thus make a good initial guess of all the lane boundaries by shifting the auto-generated ego-lane boundaries laterally by the lane width. While we do not capture complex road topologies such as splits and merges, we obtain a good springboard for the human annotation task (refer to figure \ref{fig:before_after}) .

\section{Crowdstrapping A Convolutional Neural Network for Lane Detection}

We concretely attempt to integrate the experience of people to crowdstrap learning systems for autonomous driving by using Driverseat to train and evaluate a convolutional neural network on the lane detection task. 

To acquire labelled data to train the neural network, we (a) collect data on California highways using our sensor-equipped research vehicle, (b) process the data to build 3D maps of the drives and generate an initial ground truth estimate for the lanes using automated labelling, and finally (c) use Driverseat's RoadEdit interface to have human annotators correct and label complex lane topologies.

This labeled data serves as input to the neural network, the architecture of which is detailed in \cite{brody}. The neural network's task is to predict the pixel (x, y, depth) locations of the lane boundaries given an image of the road. We create a dataset consisting of lane-labelled images by projecting the labeled 3D data from the global (x, y, z) coordinates into the camera image. The neural network is finally trained on a large subset of this data, while the rest of the data is used for hold-out validation.

To evaluate the performance of the neural network, we compare lane predictions against ground-truth labels to compute precision and recall across a range of depths. We then use Driverseat's TagEval framework to evaluate network performance in a variety of scenarios. Annotators tag road segments in the holdout set with interesting road, weather and lighting scenarios. Performance of the network is then evaluated specifically on images corresponding those scenarios, and their results quantitatively and qualitatively analyzed. We can hence focus on collecting more data for the scenarios which challenge the network, and make the system robust for the road ahead.

\section{Results}

We evaluate the performance of the convolutional neural network on the lane detection task in an array of different scenarios tagged with Driverseat's TagEval interface, covering (a) left and right off-ramps, (b) road curves, (c) shadows, and (d) pavement changes (when the road changes texture/color).

\begin{figure}[H]
\centering
\subfigure[Occasional left-side off-ramps challenge network detection of the leftmost lanes.]{
\includegraphics[width=0.45\textwidth]{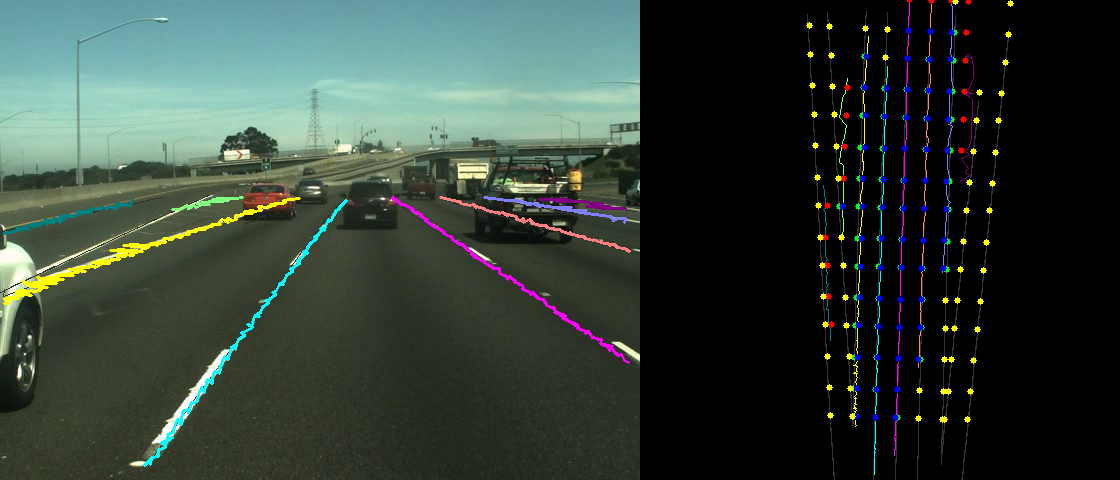}
}
\subfigure[Labels from TagEval reveal that the network is weak at capturing lanes in scenarios with off-ramps. However, note that on the left, the network is able to make lane detections even in the presence of the occluding truck.]{
\includegraphics[width=0.45\textwidth]{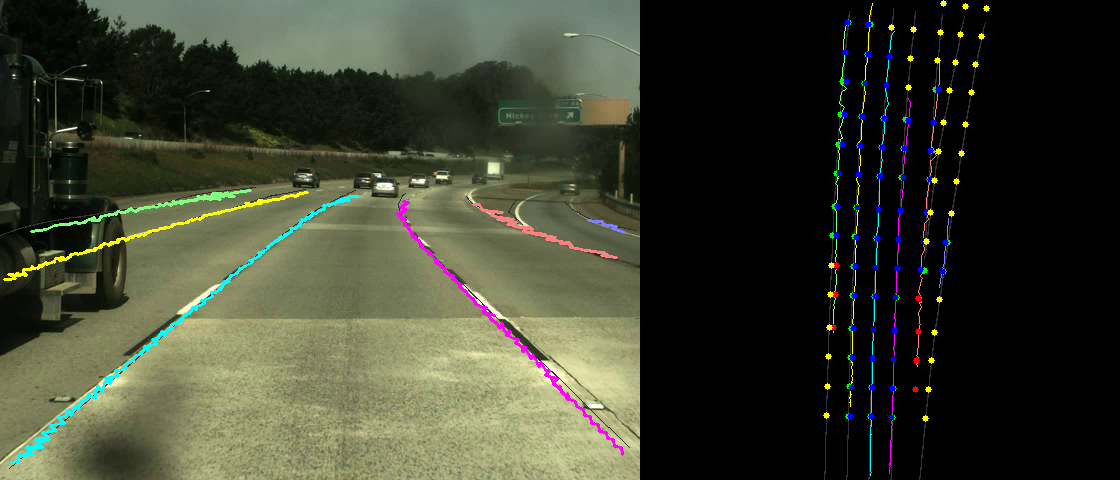}
}
\subfigure[Shadows continue to pose a natural challenge for lane detection systems.]{
\includegraphics[width=0.45\textwidth]{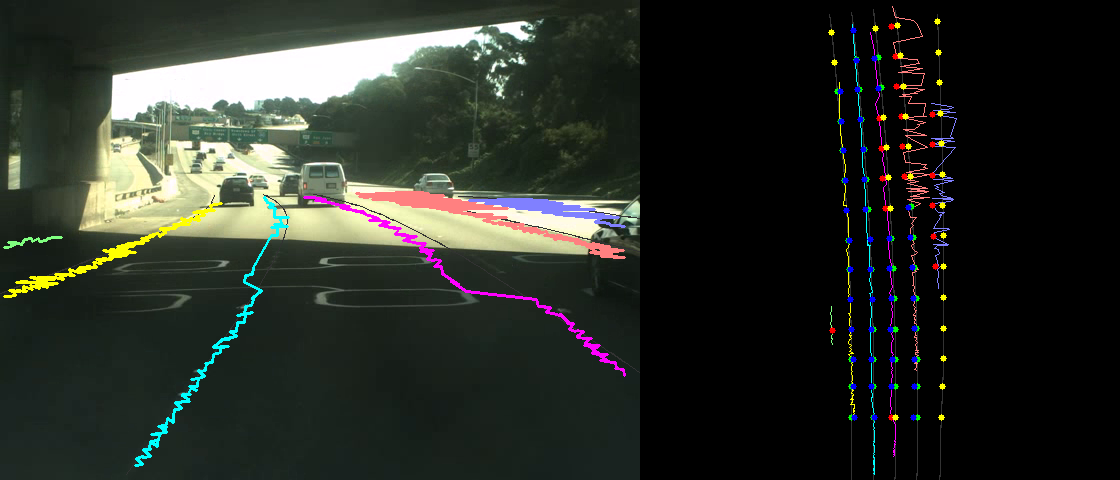}
}
\subfigure[Curves on the road, which we hypothesized would hurt the neural network's performance, are handled well.]{
\includegraphics[width=0.45\textwidth]{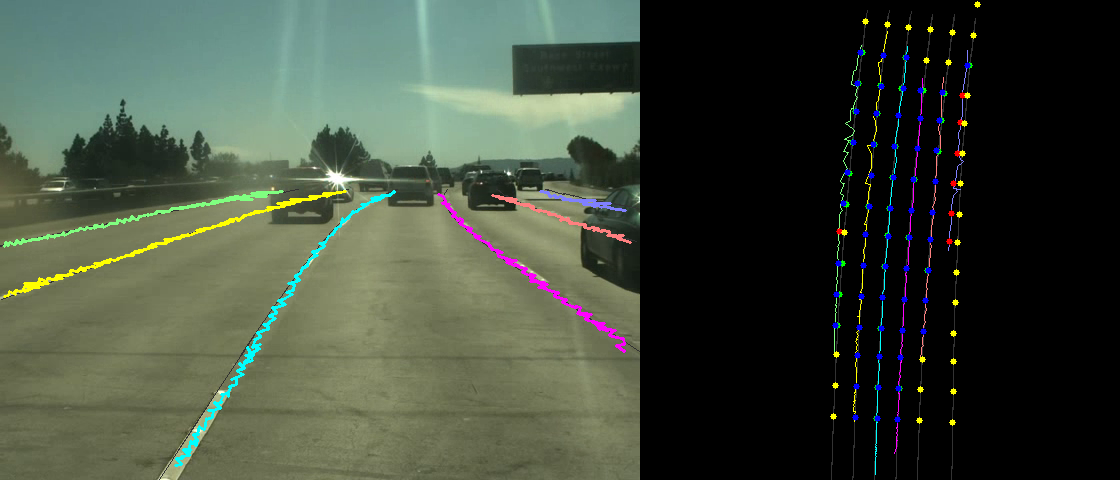}
}
\caption{Example of neural network outputs on a slew of different scenarios.}
\label{fig:example_conditions}
\end{figure}

Qualitative evaluation of network performance on these scenarios reveal strengths and difficulties for the network (see figure  \ref{fig:example_conditions}). On one hand, (a) off-ramps are challenging for the network, and emerging lanes are often missed, (b) shadows and pavement-changes are modestly difficult too, and the network becomes susceptible to misinterpreting reflective patches on the road with lane markings. On the other hand, the network is (a) robust to curves on highways, and (b) is able to detect lane boundaries even in the presence of occluding vehicles.

\begin{figure}[t]
\centering
\subfigure[ego-lane left boundary]{
\includegraphics[width=.225\textwidth]{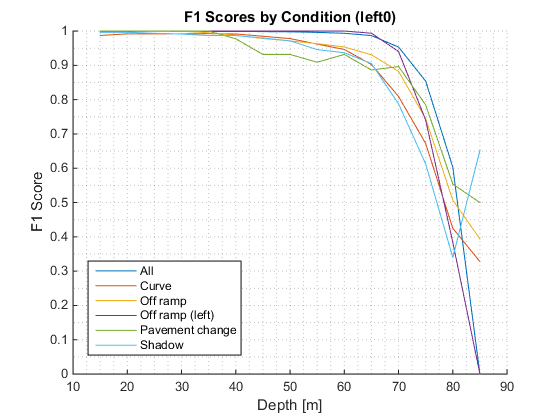}
}
\subfigure[ego-lane right boundary]{
\includegraphics[width=.225\textwidth]{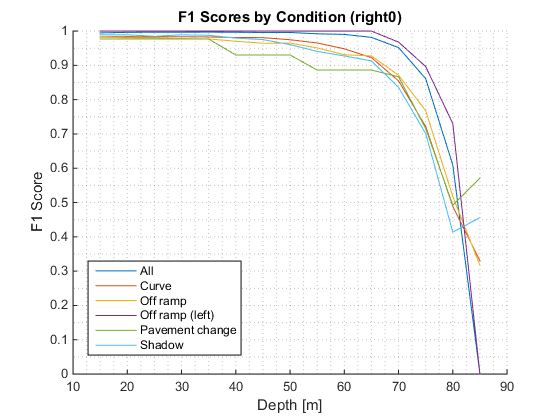}
}

\subfigure[left-adjacent lane]{
\includegraphics[width=.225\textwidth]{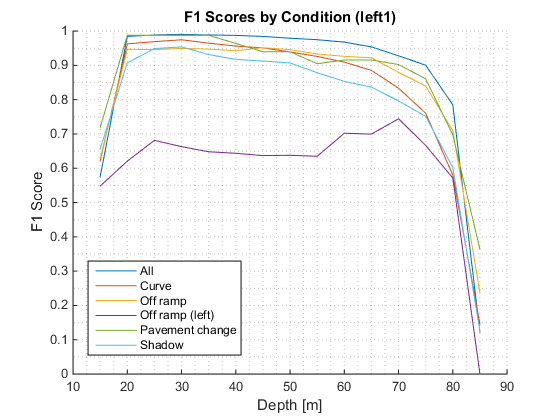}
}
\subfigure[right-adjacent lane]{
\includegraphics[width=.225\textwidth]{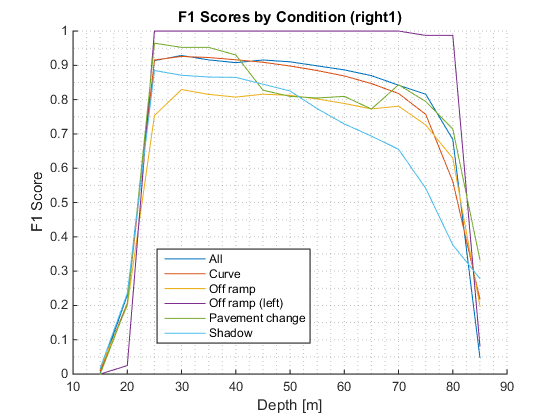}
}

\caption{Quantitative evaluation of performance in a diverse range of scenarios.}
\label{fig:quantitative_conditions}
\end{figure}

Quantitative evaluation gives further insight (see figure \ref{fig:quantitative_conditions}). For ego-lane boundary detection, we note that (a) performance on road-curves, though comparable with general performance, declines rapidly with distance (b) shadows and pavement changes are the most challenging conditions for ego-lane detection. Graphing performances on the lanes adjacent to the ego-lane on either side, we identify (a) off-ramps injure detection performance on the splitting lane, and (b) shadows maintain to be a challenging scenario.

\newpage
\section{Conclusions}
The work presents Driverseat, a system that embeds crowds around, or \textit{crowdstraps}, learning systems. In training of learning systems, we demonstrate how Driverseat uses crowd contributions to annotate labels that are hard to auto-generate. In the evaluation of learning systems, we demonstrate how Driverseat can leverage human intelligence to pinpoint scenario-specific weaknesses. To further motivate how human computation can be combined with machine computation, we demonstrate how automation can be used to provide initial estimates of ground-truth for the human annotation task. We conclude by applying all these techniques to the concrete task of lane detection in autonomous driving, a salient feature for the next generation of autonomous vehicles. We evaluate the performance of the network on a variety of scenarios, some of which are handled better than others.

As our learning systems continue to get better and more advanced, new challenges and opportunities arise. As we apply learning systems to increasingly complex problems, we need to explore more sophisticated strategies for combining human and computer intelligence. In this work, we have shown how we can integrate people's knowledge and experience on the roads to ``teach'' machines to drive, en route to our goal of making roads accident-free. Driverseat points to a future in which artificial intelligence works hand-in-hand with human intelligence to solve the most complex of problems.

\section{Acknowledgements}
We would like to thank Professor Michael Bernstein for helpful conversations about this work. We would also like to thank Mary McDevitt her support in the writing of this paper.


\bibliography{example_paper}
\bibliographystyle{icml2015}
\end{document}